\documentclass{ws-procs9x6}

\begin{document}

\title{INTERSTELLAR GAMMA RAYS AND COSMIC RAYS: NEW INSIGHTS FROM FERMI-LAT AND INTEGRAL \\
}

\author{A. W. STRONG$^*$, on behalf of the Fermi-LAT Collaboration }

\address{Max-Planck-Institute f\"ur extraterrestrische Physik\\
Garching, Germany\\
$^*$E-mail: aws@mpe.mpg.de\\
}
%www.university\_name.edu}

%\author{A. N. AUTHOR}

%\address{Group, Laboratory, Street,\\
%City, State ZIP/Zone, Country\\
%E-mail: an\_author@laboratory.com}

\begin{abstract}
In over  two years of operation Fermi-LAT has revolutionized our knowledge of the gamma-ray sky.
Interstellar gamma rays are part of this new era and allow unprecedented tests for models of cosmic rays in the Galaxy.
The extension to lower energies with INTEGRAL/SPI data is also evolving.
The global multiwavelength luminosity of the Milky Way has been derived, with implications for 
the Galactic energy balance and the radio-FIR correlation.
\end{abstract}


\bodymatter

%%%%%%%%%%%%%%%%%%%%%%%%%%%%%%%%%%%%%%%%%%%%%%%%%%%%%%
\section{Introduction}\label{aba:sec1}

In more than two years of operation so far, Fermi-LAT has revolutionized our knowledge of the gamma-ray sky.
Interstellar gamma rays are part of this new era and allow unprecedented tests for models of cosmic rays (CR) in the Galaxy.
The extension to lower energies with INTEGRAL/SPI data is also making significant advances.
 There are also new results on the global multiwavelength luminosity of the Galaxy, 
with implications for the energy balance and the radio-FIR correlation.
For a  pre-Fermi review of CR and gamma rays see Ref.~\refcite{2007ARNPS..57..285S}.

%%%%%%%%%%%%%%%%%%%%%%%%%%%%%%%%%%%%%%%%%%%%%%%%%%%%%%%%%%%%%%%%%%
\section{Fermi results on interstellar emission}\label{aba:sec2}

First I briefly summarize the main  Ferm-LAT results on interstellar gamma rays so far.
The first major result\cite{2009PhRvL.103y1101A}
 was to show that the `GeV excess' above the expected interstellar emission was evidently an EGRET instrumental effect;
instead the Fermi-LAT spectrum at intermediate latitudes (i.e. emission in the our local neighbourhood of the Galaxy) agrees rather well with that expected from locally-measured CR interacting with atomic and molecular hydrogen, plus inverse Compton emission.
A second result\cite{LAT:HIEmissivity} was a detailed study of the local gamma-ray emission from CR interacting with atomic hydrogen using a correlation analysis ; again good agreement with the expected spectrum from CR interactions was found.
Thirdly \cite{2010ApJ...710..133A} %Abdo etal 2nd quadrant
  a study of the 2nd Galactic quadrant - probing the outer Galaxy - allowed determination of the H$_2$-to-CO  factor (X$_{CO}$), confirming the increase with Galactocentric radius\cite{2004A&A...422L..47S}
  and finding a rather lower local value than previously obtained with EGRET data.
 This analysis showed a surprisingly large level of emission in the outer Galaxy. This has been confirmed by a new study of the 3rd quadrant\cite{2011_3rd_quadrant}~.
 The outer galaxy emission is greater than expected if CR sources follow supernova remnants as traced by e.g. pulsars. CR propagation in a larger halo (10 kpc height) can help to reproduce this, but
possibly large amounts of gas not traced by 21-cm atomic hydrogen  and CO (tracer of H$_2$) surveys are present in the outer Galaxy\cite{2002ApJ...579..270P}~, and this could help explain this emission.
Another aspect which requires good modelling of the Galactic emission is the determination of the extragalactic gamma-ray background\cite{EGBpaper} .

Meanwhile gamma-ray emission has been detected in five (non-AGN) extragalactic systems : LMC\cite{2010A&A...512A...7A} , SMC\cite{2010A&A...SMC}~,
 M31  (with an upper limit for M33)\cite{2010A&A...M31}  and the starburst galaxies M82 and NGC253\cite{2010ApJ...709L.152A} .
 Since this emission is presumably mainly interstellar as in the Galaxy, the combination of these results throws light on  CR in a wider context. 
These results are discussed by P. Martin  (this conference).

%%%%%%%%%%%%%%%%%%%%%%%%%%%%%%%%%%%%%%%%%%%%%%%%%%% 

\section{Comparison of large-scale Galactic emission  models with Fermi-LAT data}\label{aba:sec1}

The data are from the first year of Fermi-LAT observations, using a gamma-ray event class optimized for diffuse emission with low background\cite{EGBpaper} .
Counts and exposure were  converted to HEALPIX for processing.

The model was generated with the GALPROP\footnote{http://galprop.stanford.edu} code (see Ref.~\refcite{2007ARNPS..57..285S} and references therein).
The model includes diffusive reacceleration with a 4~kpc halo height, and details of the parameters are given in a related paper\cite{2041-8205-722-1-L58} .
It uses the locally-measured CR proton and Helium spectra, and the Fermi-LAT measurements of the electron (plus positron) spectrum, for normalization at the Sun.
For gamma-ray predictions, gas column densities from HI, CO surveys, far infrared dust emission as a tracer of additional gas,  and a model for HII, are used.
The cosmic-ray source distribution follows pulsars, which are taken as a tracer of supernova remnants, the likely  source of CR.
The radial variation of  X$_{CO}$ is taken from earlier work\cite{2004A&A...422L..47S} .
For inverse Compton, the interstellar radiation field of GALPROP\cite{2008ApJ...682..400P} is used.
The extragalactic (or at least isotropic) background is from the Fermi-LAT analysis\cite{EGBpaper}~.
Known gamma-ray sources are taken from the Fermi 1FGL catalogue\cite{FermiCatalog} and included in the model.
The model skymaps for each physical component were convolved to Fermi-LAT with the appropriate energy-dependent point-spread-function.

It is important to note that the chosen model is just illustrative, makes no claim to be unique,  nor has it been specially fitted to Fermi data (this is ongoing work to be reported elsewhere); for this reason it is referred to as an `{\it a priori} model'.
Even without such details, the quality of Fermi-LAT data allows interesting comparisons.

Figures 1 and 2 show the  spectrum of the inner Galaxy, and longitude and  latitude profiles at 1.4 GeV.
 Such a basic model reproduces the data over much of the sky within $\approx20\%$ and, remarkably, from the Galactic plane to the poles in latitude over 2 decades dynamic range.
The latter shows the importance of inverse Compton emission which dominates at high latitudes. 
The spectrum of the inner Galaxy is in reasonable agreement with the model, while there is much room for adjustment of the parameters which could improve the fit; this is the subject of forthcoming work by the Fermi-LAT collaboration. 
The latitude profile in particular can be improved with a larger halo height of about 10 kpc instead of 4 kpc (also shown in Figure 2), and this is consistent with the higher outer Galaxy emission mentioned in Section 2.
While the global agreement is good, there are significant residuals over the sky, again a subject for future detailed studies\footnote{Other studies have given indications for large-scale features not predicted by standard CR propagation models, such as the `Fermi bubbles' \cite{2010arXiv1005.5480S}~, but these will not be discussed here}.

%%%%%%%%%%%%%%%%%%%%%%%%%%%%%%%%%%%%%%%%%%%%%%%%%%%%%%%%%%%%%%
\section{Contribution from unresolved sources}

While Fermi-LAT has detected  1451 gamma-ray sources\cite{FermiCatalog} which can be accounted for in comparisons of the diffuse emission with models,
the underlying source populations will certainly contribute to the remaining emission. Current estimates put the level of this emission at 10-20\% of 
the total emission, depending on energy and direction, which corresponds to at least as much as the detected sources. This is therefore a major uncertainty affecting any comparison of diffuse models with data, especially towards the inner Galaxy. In fact, the spectrum of the inner Galaxy (Figure 1) reveals an excess at GeV energies over that expected, and this could plausibly be attributed to unresolved source populations, in particular pulsars.
A detailed study of this aspect is in preparation.
%%%%%%%%%%%%%%%%%%%%%%%%%%%%%%%%%%%%%%%%%%%%%%%%%%%%%%%%%%%%%%%%

\begin{figure}
\begin{center}
\psfig{file=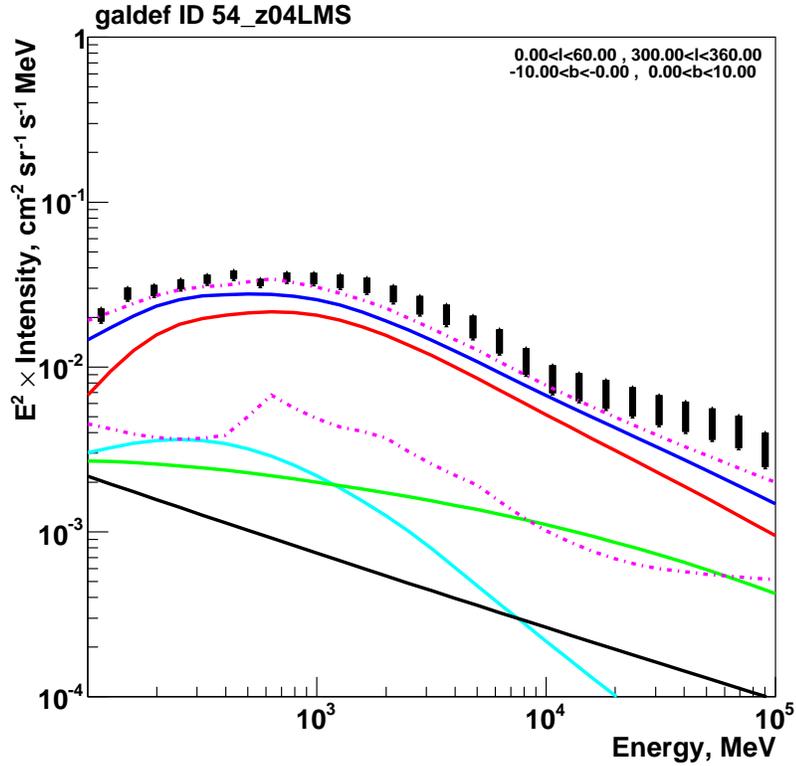,width=4.5in}
\end{center}
\caption{Spectrum of inner Galaxy  ( $300^o<l<60^o$, $|b|<10^o$).  Model from GALPROP, based on locally-measured cosmic ray spectra and a halo height 4 kpc.
Lines show the components of model; red: pion-decay, green: inverse Compton, black: extragalactic/isotropic, blue: total without sources, magenta dashed: Fermi-detected sources and total including sources. Data: black vertical bars: Fermi-LAT (PRELIMINARY).}
\label{aba:fig1}
\end{figure}

\begin{figure}
\psfig{file=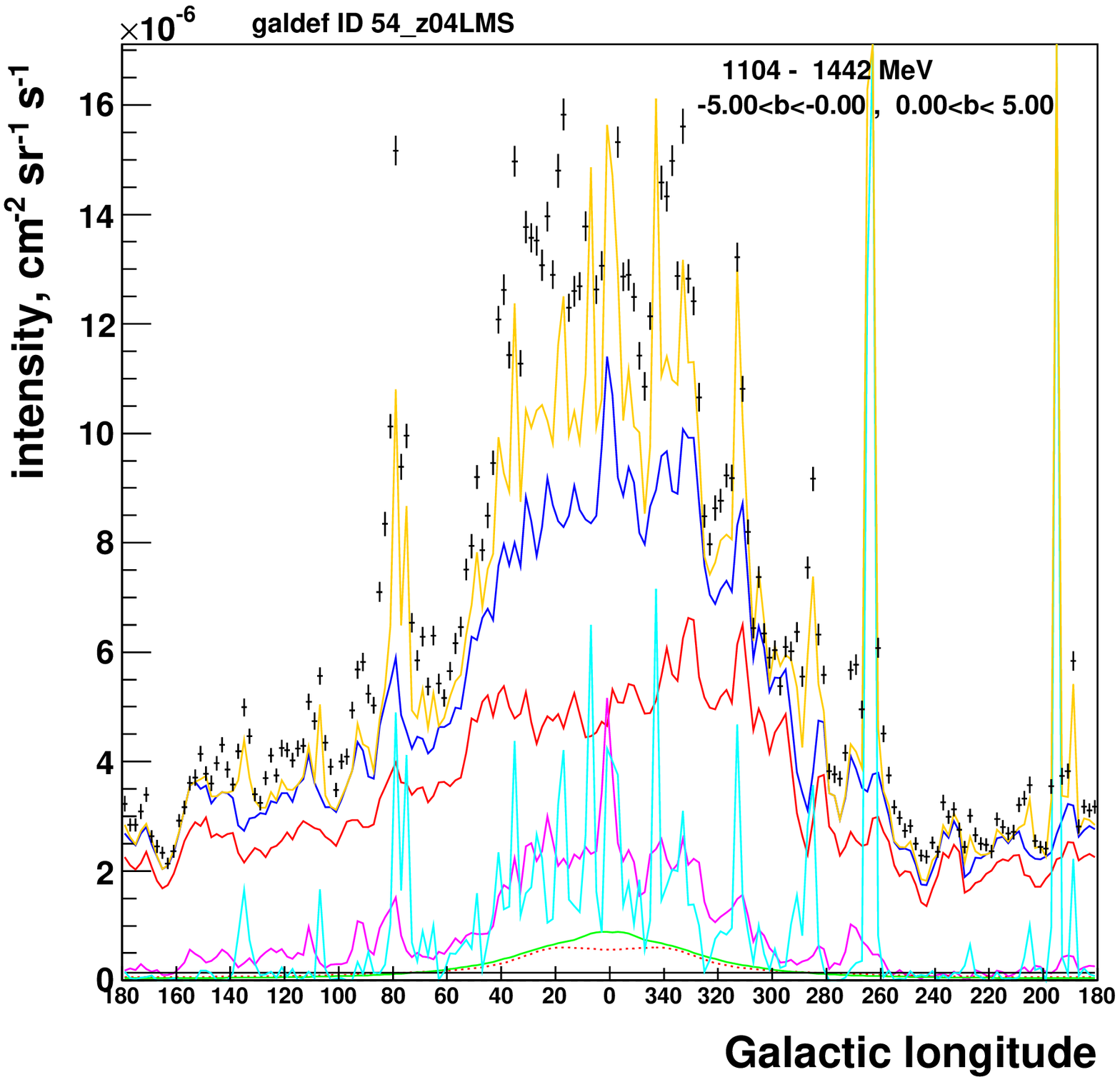,width=4.0in}
\begin{center}
\psfig{file=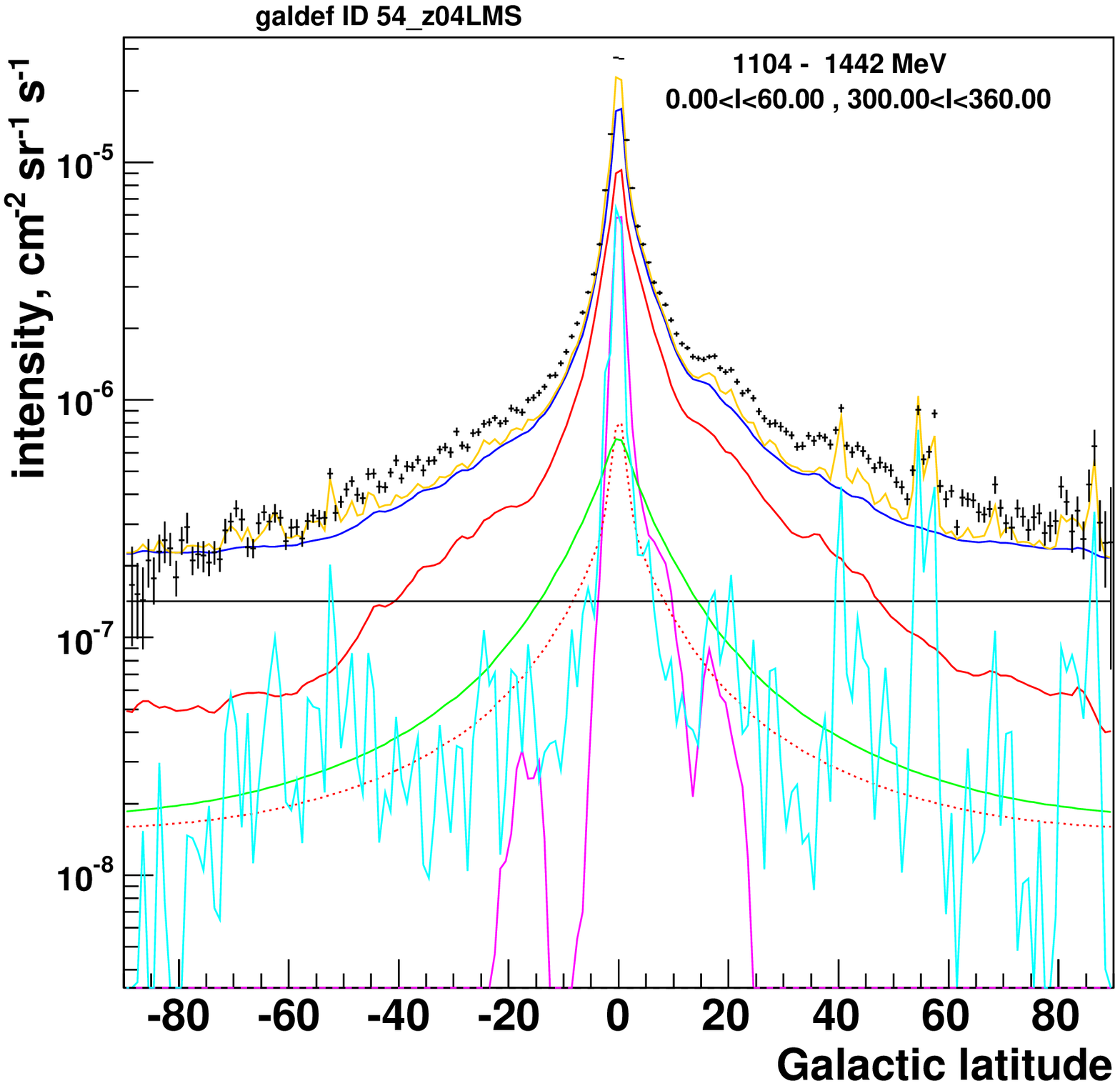,width=2.2in}
\psfig{file=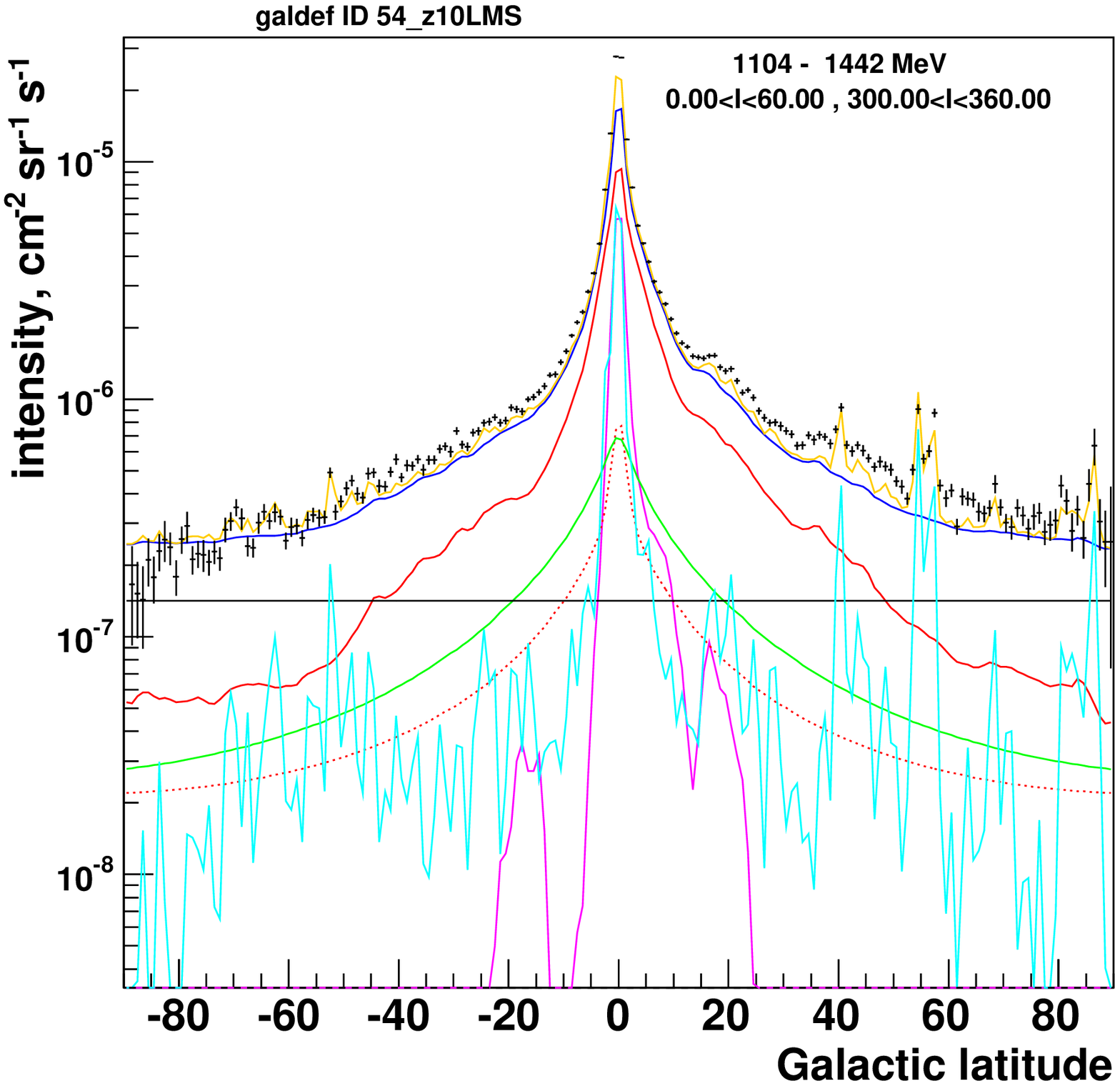,width=2.2in}
\end{center}
\caption{Longitude ($|b|<5^o$) and latitude ($300^o<l<60^o$) profiles using Fermi-LAT (PRELIMINARY) data for 1.1--1.4 GeV. Model as in Figure  1, except right latitude profile has halo height 10 kpc.
Lines show components of the model; red solid: HI, red dashed: HII,  magenta: H$_2$, green: inverse Compton,  cyan: Fermi-detected sources, black: extragalactic/isotropic, blue : total without sources, orange : total including sources.}
\label{aba:fig2}
\end{figure}

%%%%%%%%%%%%%%%%%%%%%%%%%%%%%%%%%%%%%%%%%%%%%%%%%%%%%%%
\section{Multiwavelength luminosity of the Galaxy}

Although spectral energy distributions (SEDs) are plentifully available for AGN, SNRs, pulsars and many other objects, for the Milky Way it is more difficult because of our postion inside the Galaxy. Recently\cite{2041-8205-722-1-L58} it has been possible to synthesize the multiwavelength spectrum of the Galaxy from radio to gamma rays, based on the GALPROP model which includes all CR-related processes (although not compact sources, lines etc). The model is compatible with radio, gamma-ray and CR data, but is obviously not unique; discussion of the possible range of models is given in that paper. Figure 3 shows such an SED, including the optical and infrared components as well. This is useful as the basis for estimating the gamma-ray background from normal galaxies, and for comparison with external galaxies such as M31\cite{2010A&A...M31} which are similiar to the Milky Way  (see also P. Martin, these proceedings). In addition it allows us to address the question of whether the Galaxy is a lepton calorimeter: the answer is affirmative provided both synchroton and inverse Compton emission are included.  This is useful information for the interpretation of the radio-far-infrared correlation observed for a wide range of galaxies: it provides the only case where the lepton calorimetry can be directly checked via gamma rays, synchrotron and direct measurements of cosmic ray leptons. Normally only the synchrotron component is considered in assessing calorimetry. A spin-off of this study was to show that the Milky Way lies on the radio-FIR relation defined by other galaxies.

\begin{figure}
\begin{center}
\psfig{file=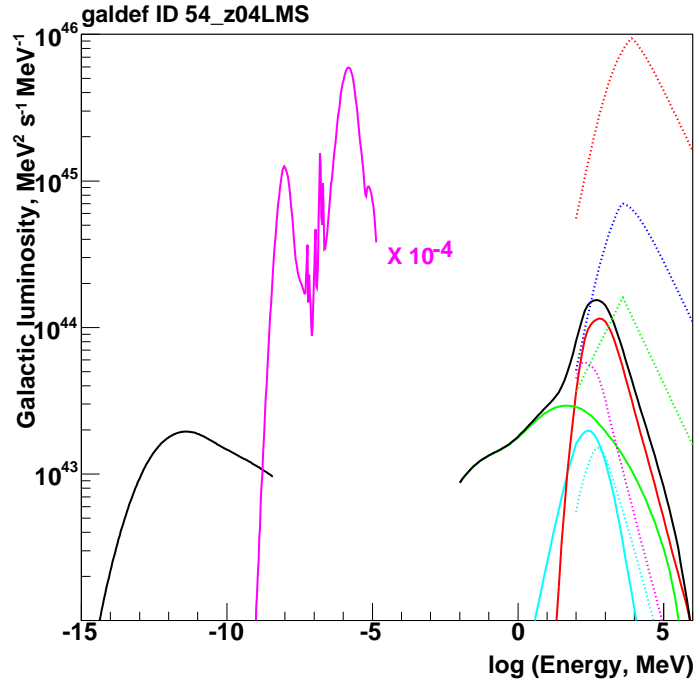,width=4.0in}
\end{center}
\caption{Luminosity spectrum of the Galaxy, from Ref.~\refcite{2041-8205-722-1-L58}.   Components: black lines: synchrotron and total gamma rays; continuous lines red: pion-decay; cyan: bremsstrahlung;  green: inverse Compton; dotted lines: cosmic rays: red: protons, blue: Helium, green: electrons, magenta: secondary positrons, cyan: secondary electrons. Magenta lines: optical and infrared, scaled by $10^{-4}$ for clarity. From Ref~\refcite{2041-8205-722-1-L58}. Reproduced by permission of the AAS.}
\label{aba:fig1}
\end{figure}

%%%%%%%%%%%%%%%%%%%%%%%%%%%%%%%%%%%%%%%%%%%%%%%%%%%%%%%%
\section{Extension to hard X-rays and low-energy gamma rays}

While Fermi-LAT has revolutionized our knowledge of the gamma-ray sky above 100 MeV, progress at lower energies is slower. In the MeV range no new data since GRO/COMPTEL have been available\footnote{The sensitivity contrast  between 1--30 MeV and $>$100 MeV is now enormous, comparing  COMPTEL with Fermi-LAT: about a factor of 50. This makes it all the more regrettable that no new MeV instrument is confirmed to fly in the coming decade. The MeV sky will remain {\it cielo incognito} for some time to come}.
 The SPI instrument on the INTEGRAL satellite has improved the situation with a spectrum of diffuse emission from 20 keV to 1 MeV\cite{2008ApJ...679.1315B,2008ApJ...682..400P} and more recently to 2 MeV (Bouchet et al. in preparation).
Figure 4 shows a combination of Fermi-LAT, COMPTEL and INTEGRAL/SPI data for the inner Galaxy. Of particular interest is the fact that inverse Compton emission is the main process in the range 100 keV to a few MeV, and accounts well for the observed power-law continuum spectrum there. Note that the data shown includes other diffuse components like annihilation positronium and 511 keV, and $^{26}$Al which are not included in the model, while below 100 keV, hard X-ray sources dominate the emission\cite{2009Natur.458.1142R} . This figure illustrates the importance of extending the observations to lower as well as higher energies. The excess in the MeV range observed by COMPTEL has been known for a long time and is plausibly attributable to discrete sources - however our lack of knowledge of the MeV sky impedes a more detailed discussion. 
\begin{figure}
\begin{center}
\psfig{file=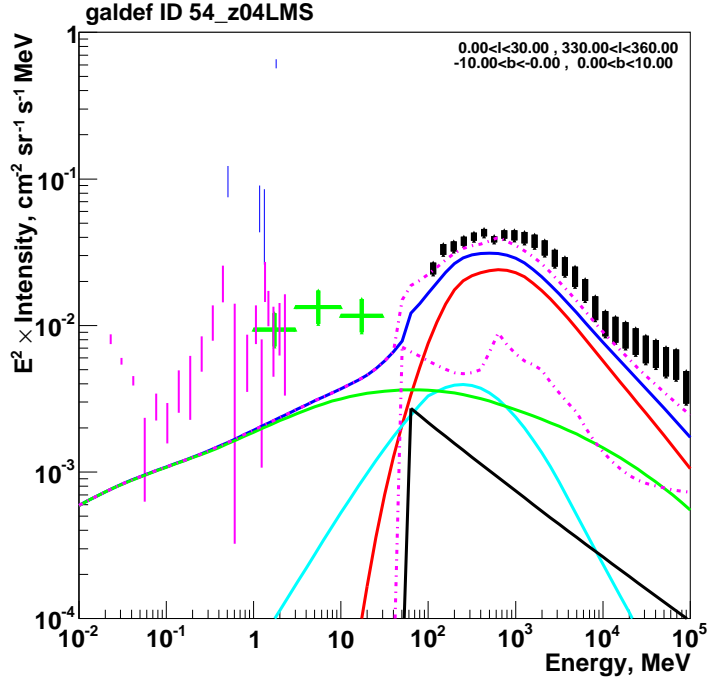,width=4.0in}
\end{center}
\caption{Spectrum of inner Galaxy ( $330^o<l<30^o$, $|b|<10^o$). Data from INTEGRAL/SPI (L. Bouchet, private communication), COMPTEL and Fermi-LAT. Model  and components coding as in Figure 1.  Data: black vertical bars: Fermi-LAT (PRELIMINARY), green crosses: COMPTEL, magenta and blue vertical bars: INTEGRAL/SPI (magenta: continuum, blue: line emission ). The longitude range is less than in Figure 1 to acommodate the INTEGRAL/SPI measurement. The model below 100 MeV does not include positronium, line emission or point sources.}
\label{aba:fig1}
\end{figure}

\section{Acknowledgements}
The $Fermi$ LAT Collaboration acknowledges support from a number of agencies and institutes for both development and the operation of the LAT as well as scientific data analysis. These include NASA and DOE in the United States, CEA/Irfu and IN2P3/CNRS in France, ASI and INFN in Italy, MEXT, KEK, and JAXA in Japan, and the K.~A.~Wallenberg Foundation, the Swedish Research Council and the National Space Board in Sweden. Additional support from INAF in Italy and CNES in France for science analysis during the operations phase is also gratefully acknowledged.

%%%%%%%%%%%%%%%%%%%%%%%%%%%%%%%%%%%%%%%%%%%%%%%%%%%%%%%%%%%%%%%%%%%%%%%%%%%%%
%\section{References}
%References are to be listed in the order cited in the text in Arabic
%numerals. \btex\ users, please use our bibliography style file
%\verb|ws-procs9x6.bst| for references. Non \btex\ users can list
%down their references in the following pattern.

%\begin{verbatim}
%\begin{thebibliography}{9}
%\bibitem{jarl88} C. Jarlskog, in {\it CP Violation} (World
%        Scientific, Singapore, 1988).

%\bibitem{lamp94} L. Lamport, {\it \LaTeX, A Document
%        Preparation System}, 2nd edition (Addison-Wesley,
%        Reading, Massachusetts, 1994).

%\bibitem{ams04} \AmS-\LaTeX{} Version 2 User's Guide (American
%        Mathematical Society, Providence, 2004).

%\bibitem{best03} B.~W. Bestbury, {\em J. Phys. A} {\bf 36},
%        1947 (2003).
%\end{thebibliography}
%\end{verbatim}

% defined these since not recognized by ws-procs9x6 bibtex macro
\def\apj{The Astrophysical Journal}
\def\apjl{The Astrophysical Journal Letters}
\def\apjs{The Astrophysical Journal Supplement Series}
\def\aap{Astronomy and Astrophysics}
\def\nat{Nature}

\bibliographystyle{ws-procs9x6}
\bibliography{FermiDiffusePaper2,extragalactic,strong,villaolmo}

\end{document}